\documentclass[3p,times,twocolumn]{elsarticle}

\usepackage{ecrc}

\volume{00}

\firstpage{1}

\journalname{Nuclear Physics B Proceedings Supplement}

\runauth{G. Morlino}

\jid{nuphbp}

\jnltitlelogo{Nuclear Physics B Proceedings Supplement}

\usepackage{amssymb}

\usepackage[figuresright]{rotating}

\begin{document}

\begin{frontmatter}



\dochead{}

\title{Using optical lines to study particle acceleration at supernova remnants}


\author[1,2]{Giovanni Morlino}

\address[1]{APC, AstroParticule et Cosmologie, Universit\'e Paris Diderot, CNRS/IN2P3, CEA/Irfu, Observatoire de Paris, Sorbonne Paris Cit\'e, 10, rue Alice Domon et L\'eonie Duquet, F-75205 Paris Cedex 13, France.}
\address[2]{Department of Physics \& Astronomy, Purdue University, 525 Northwestern Avenue, West Lafayette, IN 47907-2036, USA.}

\begin{abstract}
The shocks of several young supernova remnants (SNR) are often associated with very thin optical filaments dominated by Balmer emission resulting from charge-exchange and collisional excitation between neutral Hydrogen from the interstellar medium and shocked protons and electrons. Optical lines are a direct probe of the conditions at the shock, in particular the width of the narrow and broad components reflect the temperature upstream and downstream of the shock, respectively. When the shock accelerate efficiently non-thermal particles, the shock structure changes producing anomalous Balmer lines and it is possible to use their line shape and their spatial profile to check the efficiency of SNR shocks in accelerating cosmic rays. Here we illustrate the kinetic theory of shock acceleration in presence of neutrals with some applications to young SNRs. We show that in three cases (RCW 86, SNR 0509-67.5 and Tycho) anomalous Balmer lines can be explained assuming that a fraction of $\sim10\%$ of the total shock kinetic energy is converted into not thermal particles, while in one single case, the northwestern part of SN 1006, there is no evidence of efficient acceleration.
\end{abstract}


\end{frontmatter}


\section{Introduction}
\label{sec:intro}
In the context of supernova remnant (SNR) paradigm for the origin of cosmic rays (CR), particle acceleration takes place at shocks associated with the supernova explosion, and it is described by the non linear theory of diffusive shock acceleration (see \cite{Malkov01} for a review). Energy and momentum conservation at the shock in the presence of accelerated particles leads to two straightforward conclusions: 1) since part of the energy is channeled into particle acceleration, the thermal energy (hence the temperature) of the downstream gas is expected to be lower than in the absence of CRs; 2) the dynamical reaction of CRs induces the formation of a precursor upstream of the shock, which results in a deviation of the CR spectrum from a simple power-law behavior. Optical spectra observed from so called Balmer dominated shocks can be used to test these predictions. In fact some young SNR shocks emit optical lines mainly consisting of Balmer H$\alpha$ emission. 

The first detection of bright H$\alpha$ filaments around the remnants of Kepler, Tycho and the Cygnus Loop was reported by \cite{Minkowski56}. A peculiarity of this emission is the weakness of forbidden metal lines which implies an high temperature of the emitting region so that radiative cooling and recombination are unimportant. The interpretation of such optical emission remained a mystery up to the seminal works of \cite{ChevRay78,Chevalier80} who proposed that it can be produced by shocks propagating thorough a partially neutral gas. Their model was able to explain the intensity, spectrum and width of the filaments observed in Tycho's SNR, including the weakness of the forbidden metal lines. A peculiarity of Balmer dominated shocks, firstly reported by \cite{Chevalier80} for the Tycho's SNR, is that the H$\alpha$ line is formed by two distinct components, a narrow line with a FWHM of few tens km/s and a broad line with a FWHM of the order of the shock speed. Similar optical profiles are now observed from a bunch of young SNRs both in the Galaxy and in the Large Magellanic Cloud (for a review see \cite{Heng10}).
 
SNR shocks are collisionless and when they propagates in partially ionized medium, only ions are heated up and slowed down, while neutral atoms are unaffected to first approximation. However, when a velocity difference is established between ions and neutrals in the downstream of the shock, the processes of charge exchange (CE) and ionization are activated and this explain the existence of two distinct lines: the narrow line is emitted by direct excitation of neutral hydrogen after entering the shock front while the broad line results from the excitation of hot hydrogen population produced by CE of cold hydrogen with hot shocked protons. As a consequence, optical lines are a direct probe of the conditions at the shock, in particular the width of the narrow and broad components reflect the temperature upstream and downstream of the shock, respectively. Now, as already pointed out, if the shock accelerate particles efficiently, the shock's structure will be modified, altering the plasma temperature and inducing the formation of a CR precursor. Hence, Balmer emission could be used to provide an indirect measurement of the CR acceleration efficiency and even to gather information on the CR induced precursor.

The first clue that Balmer emission could provide evidence for the presence of accelerated particles was put forward  as a possible way to explain the anomalous width of narrow Balmer lines reported for the first time by \cite{Smith94} and \cite{Hester94}: FWHM ranging from 30 to 50 km s$^{-1}$ was detected for four SNRs in the LMC and for the Cygnus Loop, implying a pre-shock temperature around 25,000-50,000 K. Values in the same range have been reported afterwards for other SNRs (see, e.g.\cite{Sollerman03}). If this were the ISM equilibrium temperature there would be no atomic hydrogen, implying that the pre-shock hydrogen is heated by some form of shock precursor in a region that is sufficiently thin so that collisional ionization equilibrium cannot be established before the shock. Several explanations for this anomaly were proposed but only two of them was considered realistic: 1) the neutral-induced precursor and 2) the CR-induced precursor. 

Let us comment the former possibility first. When fast, cold neutrals undergo CE interactions with the slower hot ions downstream of the shock, some fraction of the resulting hot neutrals can cross the shock and move upstream. The relative velocity between these hot neutrals and the upstream ions triggers the onset of CE and ionization interactions that lead to the heating and slowing down of the ionized component of the upstream fluid. The system then tends to develop a {\it neutral-induced} shock precursor, in which the fluid velocity gradually decreases, and even more important, the temperature of ions increases as a result of the energy and momentum deposition of returning neutrals. A first attempt at investigating the broadening of the narrow line component induced by the neutral precursor was made by \cite{LimRaga96}, using a simplified Boltzmann equation for neutrals, but their calculation does not show any appreciable change of the narrow line width. This conclusion was confirmed by \cite{BlasiMor12,MorBla12}, using a fully kinetic approach able to describe the interaction between neutrals and ions in a more accurate way. The physical reason is that the ionization length-scale of returning hot neutrals in the upstream is always smaller than the CE length-scale of incoming cold neutrals. Interestingly enough, \cite{MorBla12} showed that the neutral precursor could produce a different signature, namely the presence of a third intermediate Balmer line due to hydrogen atoms that undergone charge exchange with warm protons in the neutral precursor.

The second and more promising possibilities to explain the anomalous width of narrow lines requires efficient particle acceleration which leads to the formation of a {\it CR-induced} precursor, where ionized plasma is heated before crossing the shock. If the precursor is large enough, CE can occur upstream leading to a broader narrow Balmer line. The first attempt to model this scenario was done by \cite{Wagner08} using a two-fluid approach to treat ions and CRs but neglecting the dynamical role of neutrals.  A different model was proposed by \cite{Raymond11} where momentum and energy transfer between ions and neutrals is included, but the profile of the CR-precursor is assumed {\it a-priori} . Both works concluded that the observed width of 30-50 km s$^{-1}$ can be explained using a low CR acceleration efficiency. 

From the theoretical point of view, the main difficulty in describing the structure of a collisionless shock propagating in a partially ionized medium is that neutrals have no time to reach thermalization and cannot be treated as a fluid. Steps forward in relaxing the fluid assumption have been made by \cite{HengMcCray07} and \cite{vanAdelsberg08}, even if these works neglect the modification induced by neutrals upstream of the shock. 
A more reliable interpretation of Balmer line profile requires an accurate description of the CR acceleration process where the mutual interplay between CRs, neutrals, ionized plasma and magnetic turbulence is simultaneously taken into account. Such an approach has been developed by \cite{MorBla13} using a semi-analytical technique. This work showed that the main physical effect able to broaden the narrow line is the damping of magnetic turbulence in the CR precursor while the adiabatic compression alone is ineffective. Hence the observed widths are compatible also with large acceleration efficiency provided the right level of magnetic damping.  

Efficient CR acceleration can also affect the width of broad lines. In fact, when a sizable fraction of the ram pressure is channeled into non-thermal particles, the plasma temperature behind the shock is expected to be lower, and this should reflect in a narrower width of the broad H$\alpha$ line. Remarkably, there are clues of this phenomenon in two different remnants, RCW 86 \cite{Helder13, RCW86}, and SNR 0509-67.5 in the LMC \cite{Helder10, MorSNR0509}. In both cases the measured FWHM of the broad lines is compatible with theoretical predictions only assuming fast electron-proton equilibration downstream of the shock, a conclusion which seems to be at odds with both theoretical models and observations \cite{Rakowsky05}.

\section{The kinetic approach}
\label{sec:model}
Here we summarize the kinetic model for shock particle acceleration in presence of neutrals developed in \cite{BlasiMor12, MorBla12, MorBla13}. 
We consider a stationary system with a plane-parallel shock wave propagating in a partially ionized proton-electron plasma with velocity $V_{sh}$ along the $z$ direction. The fraction of neutral hydrogen is fixed at upstream infinity where ions and neutrals are assumed to be in thermal equilibrium with each other. The shock structure is determined by the interaction of CRs and neutrals with the background plasma. Both CRs and neutrals profoundly change the shock structure, especially upstream where both create a precursor: the CR-induced precursor reflects the diffusion properties of accelerated particles and has a typical spatial scale of the order of the diffusion length of the highest energy particles. The neutral-induced precursor develops on a spatial scale comparable with a few interaction lengths of the dominant process between CE and ionization. The downstream region is also affected by the presence of both CRs and neutrals and the velocity gradients that arise from ionization have a direct influence on the spectrum of accelerated particles.
A self consistent description of shock particle acceleration in presence of neutral Hydrogen, requires to consider four mutually interacting components: thermal particles (protons and electrons), neutrals (hydrogen), accelerated protons (CRs) and turbulent magnetic field. We neglect the presence of helium and heavier chemical elements.
The interaction terms make the system highly non linear and the solution is found using a iterative scheme similar to the one we introduced in some previous works \cite{Blasi02, AmatoBlasi05, BlasiMor12}. 

Let us start with the description of neutrals. The main difficulty arises from the fact that neutrals cannot be described as a fluid, because in the downstream the collisional ionization length is smaller than the equilibration length. Hence neutrals are described kinetically, using  the stationary Boltzmann equation to calculate the evolution of the velocity distribution function, $f_N(\vec v,z)$,
\begin{equation} \label{eq:vlasov}
v_z \frac{\partial f_{N}(\vec v, z)}{\partial z} = \beta_{N} f_{i}(\vec v, z)  -
        \left[ \beta_{i} + \beta_e \right] f_{N}(\vec v, z) \,,
\end{equation}
where $z$ is the distance from the shock (which is located at the origin), $v_z$ is the velocity component along the $z$ axis and the electron and proton distribution functions, $f_i(\vec v,z)$ and $f_e(\vec v,z)$, are assumed to be Maxwellian at each position. The collisional terms, $\beta_k f_l$, describe the interaction (due to CE and/or ionization) between the species $k$ and $l$. The interaction rate $\beta_k$ is formally written as
\begin{equation} \label{eq:beta_k}
\beta_k (\vec v,z) = \int d^{3} w \, v_{rel} \, \sigma(\vec v_{rel})
                  f_{k}(\vec w,z) \,,
\end{equation}
where $v_{rel} = |\vec v- \vec w|$ and $\sigma$ is the cross section for the
relevant interaction process. More precisely, $\beta_N$ is the rate of CE of an
ion that becomes a neutral, $\beta_i$ is the rate of CE plus ionization of a
neutral due to collisions with protons, while $\beta_e$ is the ionization rate
of neutrals due to collisions with electrons. A full description of the cross
sections used in the calculations can be found in \cite{MorBla12}.

The isotropic distribution function of CRs satisfies the following transport equation in
the reference frame of the shock:
\begin{equation} \label{eq:trasp_CR}
 \frac{\partial}{\partial z} \left[ D(z,p) \frac{\partial f}{\partial z}
 \right]
 - u \frac{\partial f}{\partial z} 
 + \frac{1}{3} \frac{d u}{d z} \, p \frac{\partial f}{\partial p} 
 + Q(z,p) = 0  \,.
\end{equation}
The $z$-axis is oriented from upstream infinity $(z=-\infty)$ to downstream
infinity $(z=+\infty)$ with the shock located at $z=0$. We assume that the
injection occurs only at the shock position and is monoenergetic at $p=p_{\rm
inj}$.
The diffusion properties of particles are described by the diffusion
coefficient $D(z,p)$. We assume Bohm diffusion in the local amplified magnetic
field:
\begin{equation} \label{eq:Diff}
 D(z,p) = \frac{1}{3} c r_L[\delta B(z)] \, ,
\end{equation}
where $r_L(\delta B)= pc/[e \delta B(z)]$ is the Larmor radius in the amplified magnetic field. The calculation of $\delta B$ is described assuming that the only turbulence which scatters particles is the one self-generated by the particles themselves through the resonant streaming instability. These waves are also damped due to several processes. In particular, when the plasma is not fully ionized, the
presence of neutrals can damp Alfv\`en waves via ion-neutral damping.  
The equation for transport of waves can be written as:
\begin{equation} \label{eq:wave_tr}
  \partial_z F_w=  u(z) \,\partial_z P_w +
   P_w \left[ \sigma_{\rm CR}(k,z)- \Gamma_{\rm TH}(k,z)   \right] \,,
\end{equation}
where $F_w(k,z)$ and $P_w(k,z)$ are, respectively, the energy flux and the pressure per unit logarithmic bandwidth of waves with wavenumber $k$. $\sigma$ is the growth rate of magnetic turbulence, while $\Gamma_{\rm TH}$ is the damping rate. For resonant wave amplification the growth rate of Alfv\'en waves is:
\begin{equation} \label{eq:sigma_CR}
 \sigma_{\rm CR}(k,x)= \frac{4\pi}{3} \frac{v_A(x)}{P_w(k,x)} \left[ 
    p^4 v(p) \frac{\partial f}{\partial x} \right]_{p=\bar p(k)} \,,
\end{equation}
where $p=\bar p(k)= eB/k m_p c$ is the resonant momentum. The damping of the waves is mainly due to non-linear Landau damping and ion-neutral damping. For the sake of simplicity here we adopt a phenomenological approach in which the damping results in a generic turbulent heating (TH) at a rate  $\Gamma_{\rm TH} = \eta_{\rm TH} \sigma_{\rm CR}$. This expression assumes that a fraction $\eta_{\rm TH}$ of the power in amplified waves is locally damped and results in heating of the background plasma. 

Finally we need to describe the dynamics of the background plasma which is affected by the presence of accelerated particles and by CE and ionization of neutrals. Protons and electrons in the plasma are assumed to share the same local density, $\rho_i(z)=\rho_e(z)$, but not necessarily the same temperature, i.e., $T_{i}(z)$ may be different from $T_{e}(z)$. The equations describing the conservation of mass, momentum and energy taking into account the interactions of the plasma fluid with CRs are:

\begin{equation} \label{eq:rh1}
 \frac{\partial}{\partial z} \left[\rho_i u_{i} + \mu_N  \right]=0 \,,
\end{equation}
\begin{equation} \label{eq:rh2} 
 \frac{\partial}{\partial z} \left[ \rho_i u_{i}^{2} + P_{g} + P_{c} + P_{w} 
        + P_{N}  \right]=0 \,,
\end{equation}
\begin{equation} \label{eq:rh3}
 \frac{\partial}{\partial z} \left[ \frac{1}{2} \rho_i u_{i}^{3} + 
  \frac{\gamma_{g} P_{g} u_{i}}{\gamma_{g}-1} + F_w + F_{N} \right]
  = -u_i \frac{\partial P_c}{\partial z} + \Gamma P_w \,.
\end{equation}
Here $\mu_N = m_H \int d^{3} v v_{\parallel} f_{N}$, $P_N = m_H \int d^{3} v v_{\parallel}^{2} f_{N}$ and $F_N = m_H/2 \int d^{3} v v_{\parallel} (v_{\parallel}^{2} + v_{\perp}^{2}) f_{N}$ are respectively the fluxes of mass, momentum and energy of neutrals along the $z$ direction (labelled as $\parallel$).  They can be easily computed once the neutral distribution function is known. $P_w$ and $F_w$ are the pressure and energy flux of waves, while $P_c$ is the CR pressure computed from the CR distribution function:
\begin{equation}
 P_c(z) = \frac{4 \pi}{3} \int dp \, p^3 v(p) f(z,p) \,.
\end{equation}
The dynamical role of electrons in the conservation equations is usually neglected due to their small mass. However, collective plasma processes could contribute to equilibrate electron and proton temperatures, at least partially. If the equilibration occurs in a very efficient manner, the electron pressure cannot be neglected and the total gas pressure needs to include both the proton and electron contributions, namely $P_{g} = P_{i} + P_{e} = P_{i}(1+\beta)$, where $\beta(z)\equiv T_e/T_i$ is the electron to proton temperature ratio and is taken here as a free parameter. While it is well established that electron-ion equilibration in the downstream might be only partial \cite{Ghavamian01,Ghavamian07},  in the presence of a precursor (either induced by the CRs or by the neutrals),  also upstream of the shock the level of equilibration becomes an unknown.

In order to solve the set of non-linear equations involving neutrals, ions, CRs and magnetic field, we adopt an iterative method that is fully described in \cite{MorBla13}. The input quantities are the values of the shock velocity and all environmental quantities at upstream infinity, where the distribution function of neutrals is assumed to be Maxwellian at the same temperature as that of ions. At the end the calculation provides the distribution function of neutrals as well as the spatial and velocity profile of Balmer emission.

\section{CR-precursor vs. neutral-precursor}
\begin{figure}
\begin{center}
{\includegraphics[width=\linewidth]{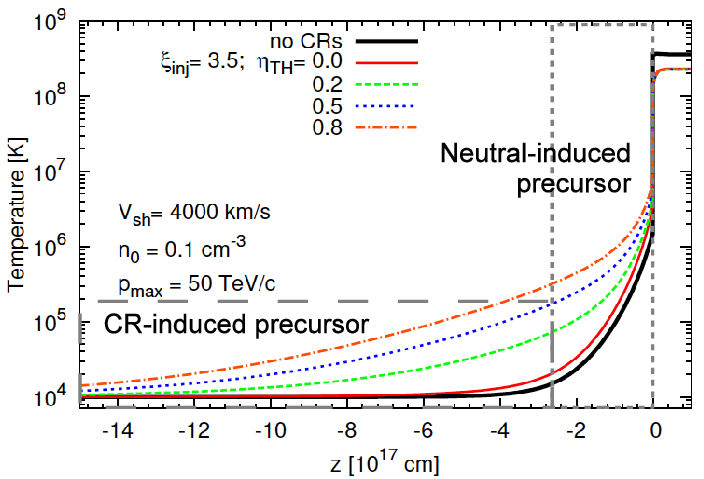}}
  \caption{Temperature of the plasma in the shock region for $V_{\rm sh}=4000\,\rm km\,s^{-1}$, $n_0= 0.1\,\rm cm^{-3}$, $h_N= 0.5$ and$B_0=10\,\mu$G. When CR acceleration is turned on, we assume $\xi_{\rm inj}=3.5$ and $p_{\rm max}=50$~TeV/c. The thick solid (black) line refers to the case with no CRs. The thin (red) solid line refers to the case with CRs but no damping of magnetic turbulence (i.e. $\eta_{\rm TH=0}$). The remaining thin curves are calculated for increasing values of $\eta_{\rm TH}$, as labelled. The dashed and dotted rectangles mark the CR and the neutral-induced precursor, respectively.} 
 \label{fig:T_precursor}
\end{center}
\end{figure}

The formalism presented above represents the first theory of particle acceleration at collisionless shocks in the presence of neutral atoms and allows us to calculate the shock structure, the spectrum of accelerated particles and the Balmer emission from the shock region. 
A good way to visualize the modification induced on the shock by the presence of both neutrals and CRs is to look at the temperature profile of ions in the shock region. In the top panel of Fig.~\ref{fig:T_precursor} we compare the temperature profile for a case with and without CRs. We chose a test case with values of parameters typical of a SNR expanding in the cold ISM, at the beginning of the Sedof-Taylor phase, i.e. $V_{\rm sh}= 4000$ km s$^{-1}$, total ISM density of 0.1 cm$^{-3}$ with 50\% of ionization fraction and temperature $T_{\rm ISM}= 10^4$ K. The black-tick solid line shows the temperature profile without CRs: the temperature increase produced ahead of the shock, on the scale of $\sim 10^{17}$ cm is entirely due to the return flux of neutrals which deposit upstream a fraction of energy subtracted from the downstream. Indeed the length-scale corresponds to the collisional ionization length of returning neutrals with upstream ions. The remaining thin curves show the case when CR acceleration is turned on. For all these curves we assume the same acceleration efficiency of 40\% (determined by the injection parameter $\xi_{\rm inj}=3.5$) and the same maximum proton energy of 50 TeV, but we change the damping efficiency of magnetic turbulence. When the damping is absent, the temperature profile is only slightly different from the case without CRs, the difference being due to the adiabatic compression produced by the CR pressure, while, when the damping is turned on, the temperature increases in the whole CR-precursor in a way proportional to the amount of damping. Notice that the typical CR-precursor length is $D(p_{\max})/V_{\rm sh} \approx 4 \times 10^{18} \delta B_{\mu \rm G}^{-1}$ cm, where $\delta B$ is the amplitude of Alfv\'en waves which resonate with particles at $p=p_{\max}$. We further notice that all cases with efficient CR acceleration have a temperature in the downstream smaller than $\sim 40\%$ with respect to the case without CRs.

\begin{figure}[t]
\begin{center}
{\includegraphics[width=1\linewidth]{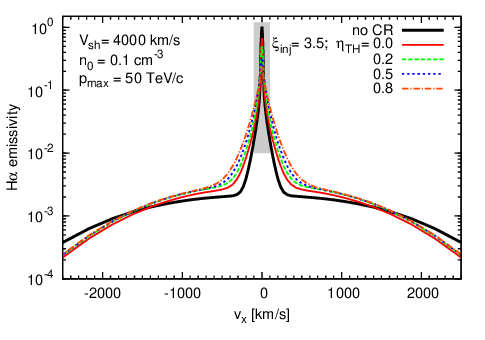}
\includegraphics[width=\linewidth]{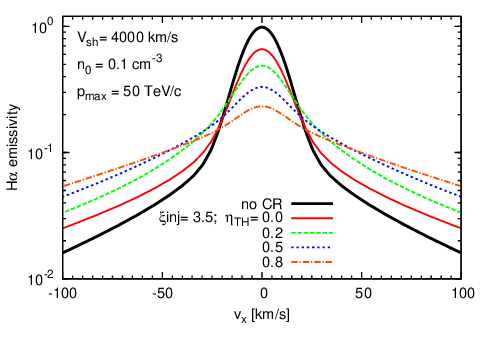}}
\caption{Volume-integrated line profile of Balmer emission for the same cases shown in Fig.~\ref{fig:T_precursor}. The lower panel shows a zoom in of the narrow line region, showed with a shadow box in the top panel. The thick solid line shows the case without CRs, while different thin lines are calculated at fixed $\xi_{\rm inj}= 3.5$ and for different values of the TH efficiency as labelled.}
\label{fig:Balmer1}
\end{center}
\end{figure}

The shape of the spatially-integrated Balmer line emission is plotted in the top panel of Fig.~\ref{fig:Balmer1}, while the bottom panel shows a zoom in the region of the narrow Balmer lines. The curves refer to the same cases as in Fig.~\ref{fig:T_precursor}, labelled as indicated. The width of the broad component of the Balmer line is appreciably reduced when CR acceleration is turned on, but the results are not sensitive to the amount of TH. On the other hand, the narrow Balmer line becomes broader when CR acceleration is efficient; the effect is more pronounced when non-negligible TH is taken into account. Hence, the distribution function of neutrals becomes broader mainly because of the scattering with a warmer ion distribution in the far precursor (i.e., where TH is more effective), rather than because of the neutral return flux, which operates only within a few CE interaction lengths from the sub-shock. 

\section{Application to young SNRs}
 \label{sec:application}
In this paragraph we compare the results of the kinetic theory for Balmer dominated shocks with few noticeable cases of young SNRs. We concentrate only on the two more direct observables  from Balmer emission, namely the width of the spatially-integrated broad and narrow lines. Other possible observables, like the spatial profile, the intermediate line or the narrow-to-broad intensity ratio could also be used but, in order to constraint the model, we need more accurate data that the ones available in the literature up to know. 
In the follow we analyze the broad Balmer emission from three remnants, SN 1006, RCW 86 and SNR 0509-65.7, while for the narrow line we show only one single case, the Tycho's remnant.

\begin{figure}
\begin{center}
{\includegraphics[width=\linewidth]{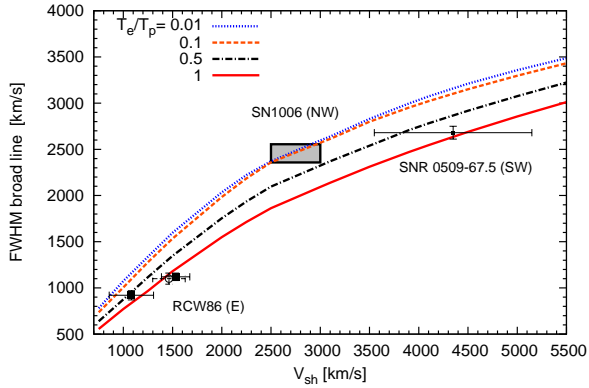}}
  \caption{Measured FWHM of the broad Balmer line as a function of the shock speed for three different remnant: RCW 86 (three different locations, one in the northeast (open circle) and two location in the southeast (filled square), data from \cite{Helder13} assuming a distance from the remnant of 2.5 kpc); northwest rim of SN 1006 \cite{Nikolic13}; southwest rim of SNR 0509-67.5 (the FWHM is taken from \cite{Helder10} while the uncertainty on $V_{\rm sh}$ is taken from the theoretical model in \cite{MorSNR0509}). Lines show the theoretical prediction without CR acceleration for different values of  electron to proton temperature ratio in the downstream \cite{MorBla13}.} 
 \label{fig:FWHM_broad}
\end{center}
\end{figure}

The width of the broad Balmer line is probably the most powerful tool to infer the presence of non-thermal particles. If the velocity of the shock is known, the width of the broad line depends mainly on two factor: the electron-to-proton temperature ratio, and the CR acceleration efficiency, while it does not depend on the total ISM density and has a weak dependence on the ionization fraction. In Fig.~\ref{fig:FWHM_broad} we show the theoretical prediction of the FWHM of the broad lines as a function of the shock speed and for different value of $T_e/T_p$, in the absence of CRs. The plot also shows the data for the three SNRs mentioned above. When measured data lie between the top and the bottom lines, they are in principle compatible with the absence of CRs, unless one is able to measure the electron temperature, breaking the degeneracy between different lines. In the following we discuss separately each single remnant.

\subsection{SN 1006}
The remnant of SN 1006 is a special case in that it presents hints of efficient particle acceleration only in the northeastern (NE) and in the southwestern (SW) regions. Here non-thermal X-ray have been detected, concentrated in thin filaments. As extensively discussed in literature, the thickness of X-ray filaments is interpreted as due to rapid synchrotron losses of electrons in a strong magnetic field ($\sim100 \mu$G), which could be produced by CR-induced amplification (if the CR acceleration is efficient). Moreover the same regions have been detected in $\gamma$-rays emission \cite{Acero10}. A further indirect indication supporting efficient CR acceleration is the distance between the contact discontinuity and the forward shock, which is smaller in the NE and SW with respect to the rest of the remnant. In fact, when a non negligible fraction of the shock kinetic energy is converted into CRs, the downstream plasma become more compressible and a reduced distance is expected.

Now, data for the FWHM of the broad Blamer line shown in Fig.~\ref{fig:FWHM_broad} (shaded box) are taken from the NW region \cite{Nikolic13} where none of the above signatures have been detected. Remarkably enough, the detected FWHM is fully compatible with theoretical expectation without CRs and assuming a low level of electron-proton equilibration ($T_e/T_p < 0.2$). 
This conclusion is further supported by the width of the narrow Balmer line which, in the NW region, is 21 km s$^{-1}$ \cite{Sollerman03}, corresponding to an ISM temperature of $10^4$ K, compatible with absence of a CR precursor. It would be extremely interesting to measure the width of Balmer lines also from the NE and the SW, but the emission in those regions is so faint that tens of hours of observation are needed to reach the required signal to noise ratio.

\subsection{RCW 86}
\begin{figure}
\begin{center}
{\includegraphics[width=\linewidth]{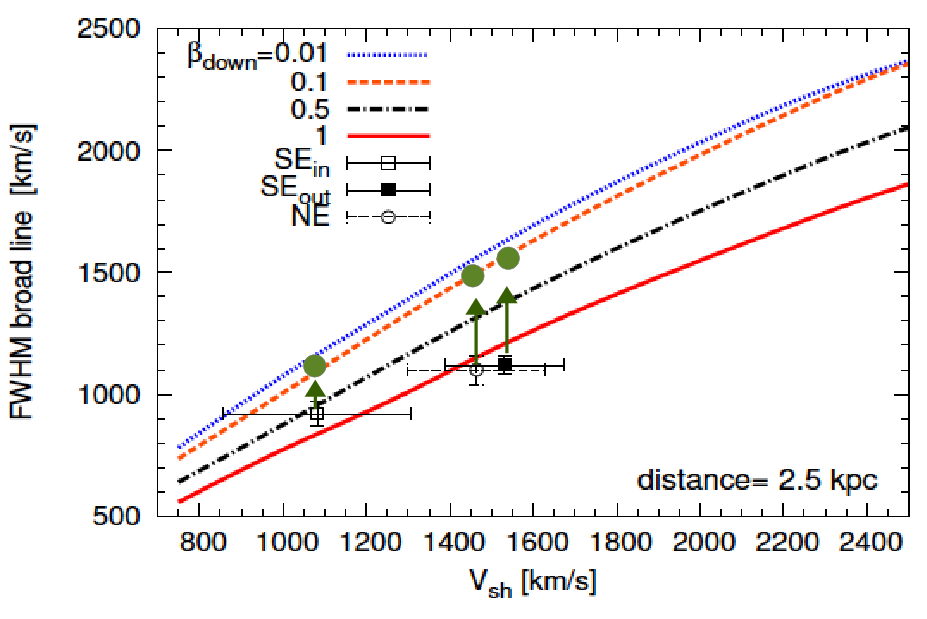}}
  \caption{FWHM measured in three different regions of the eastern part of RCW 86 by \cite{Helder11} (see their Fig. 3). Lines show the theoretical prediction without CRs as in Fig.~\ref{fig:FWHM_broad}. The filled circles show the value that the FWHM should have in these three locations if one account for the measured electron temperature and if no efficient acceleration were present.} 
 \label{fig:RCW86_FWHM}
\end{center}
\end{figure}

RCW 86 is a shell like SNR and a $\gamma$-ray source, detected both in GeV range by FermiLAT and in the TeV range by HESS. The NE part also shows non-thermal X-rays, suggesting a possible efficient production of CRs. The remnant presents many Balmer filaments around the external shell and, recently, the proper motion and the line width of these filaments have been measured \cite{Helder13}. Here we summarize the results presented in \cite{RCW86}, where the analysis has been concentrated on the eastern part of the remnant because this is the region showing the highest shock speed, hence the one with a larger probability to accelerate CRs. Because the proper motion is known, the results depends on the assumed distance from us, which is very uncertain, but generally bound between 2 and 3 kpc. The most quoted value in literature is 2.5 kpc, which will be used in the following discussion.
The three data points shown in Fig.~\ref{fig:RCW86_FWHM} represent the FWHM measured in three different regions of the eastern part of the remnant by \cite{Helder11}. The error bars in the shock speed reflect the error in the measured proper motion. The three point seem to be marginally compatible with absence of CR acceleration if we assume a full electron-proton equilibration. On the other hand this assumption is at odd with the results by \cite{Helder11}, where, using the thermal X-ray emission, the authors were able to infer the electron temperature  in the same regions where the Balmer lines have been detected. These values can be used as upper limits for the electron temperature behind the shock, because they are obtained integrating onto a large region in the downstream where Coulomb collisions can enhance the electron temperature. This information allows us to fix an upper limit to the value of $T_e/T_p$ and to break the degeneracy between different theoretical curves. We get $T_e/T_p \lesssim 0.1$ and in Fig.~\ref{fig:RCW86_FWHM} we show, with filled circles, what should be the positions of the measured FWHM if the constrain on the electron temperature is taken into account (always assuming a distance of 2.5 kpc). Hence, to explain the measured value of the FWHM,  we conclude that a fraction $\sim 20\%$ of the shock energy is converted into non-thermal particles. On the other hand, if we consider the uncertainty on the distance, the above value ranges between few per cent to 40\% for $d=2$ and 3 kpc, respectively.

An important remark is in order. The shock speed inferred for RCW 86 is relatively low, being $<2000$ km s$^{-1}$, hence we do not expect particles accelerated up to very high energies. Moreover, if the neutral fraction is $>10\%$, for such low shock speed the neutral return flux could be strong enough to affect the acceleration in the direction to make the spectrum steeper, to the point that the non-thermal energy is dominated by particles with a momentum close to the injection momentum. If this were the case, the $\gamma$-ray spectrum observed by FermiLAT and HESS could not be explained by hadronic interaction, but is more probably due to electron inverse-Compton, has suggested in \cite{Yuan14} on the basis of the hard spectrum measured by FermiLAT.

\subsection{SNR 0509-67.5}
The Balmer emission from SNR 0509-67.5 has been studied in \cite{MorSNR0509}. SNR 0509-67.5 is located in the LMC, so that its distance from the Sun is known to be $50\pm1$ kpc. At the moment there are no published measurements for the shock proper motion, but the shock speed can be inferred from evolution models taking into account the measured size of the remnant and its age, which is estimated from the light echo to be  $400\pm120$ yr. The point showed in Fig.~\ref{fig:FWHM_broad} refers to the FWHM measured in the SW rim of the remnant\cite{Helder10}, and the uncertainty in the shock speed correspond to the theoretical estimate done in \cite{MorSNR0509}. Unfortunately the big uncertainty in the determination of the shock speed prevents a firm conclusion on the CR acceleration efficiency. If the shock moves faster than $\sim4500$ km s$^{-1}$, one can conclude that particle acceleration must be taking place with an efficiency of several tens of percent. For lower shock velocity the evidence of particle acceleration becomes less clear because of the uncertainty in the electron-ion equilibrium downstream. We can speculate that if $T_e/T_p$ is as low as the value inferred in SN~1006, than the acceleration efficiency of the SW rim should be $\gtrsim 10\%$. It is worth noting that this speculation is suggested by indirect measurements of the electron-ion equilibration made in several remnants using different techniques, which point towards an inverse relation between the electron-to-ion temperature ratio and the shock speed \cite{Rakowsky05}. On the other hand none of these measurement has been done for SNR 0509-67.5 up to now.

\subsection{Tycho}
The Tycho's SNR is the best known candidate where to look for efficient CR acceleration mainly because the observed $\gamma$-ray spectrum is very steep and is difficult to explain it using leptonic processes, while pion decay produced in hadronic collisions can fit the observations very well, assuming a CR acceleration efficiency $\sim 10\%$ \cite{Tycho12}. Moreover many other observations can be simultaneously explained in the framework of non-linear shock acceleration using the same level of efficiency (namely Radio and X-ray spectrum, radio end X-ray morphology, distance between contact discontinuity and forward shock).    
Unfortunately the Balmer emission in Tycho is faint even if it is present all over around the remnant. In spite of this there is a small region of the shock, called {\it knot-g}  where the Balmer emission is enhanced, probably due to a larger ISM density in that region. These region has been analyzed by several authors and presents two interesting anomalies which could be related to efficient acceleration. 

1) A gradual increase of H$\alpha$ intensity has been measured just ahead of the shock front \cite{Lee10}. This has been interpreted as emission from the thin shock precursor ($\sim1''$ which implies a thickness of $\sim3\times 10^{16}$ cm for a distance of 3 kpc) likely due to CRs. If confirmed, this detection would represent the first direct proof of the existence of a CR precursor. On the other, hand Balmer emission from the upstream can be produced also by the neutral-induced precursor, as showed in \cite{MorBla12}, and, in order to distinguish between these two possibilities, a careful modeling of the shock is required, taking into account the the complex interaction between the CR and the neutral induced precursor. At the moment the most promising technique seems to be the kinetic theory developed in \cite{MorBla13}.

\begin{figure}[t]
\begin{center}
{\includegraphics[width=\linewidth]{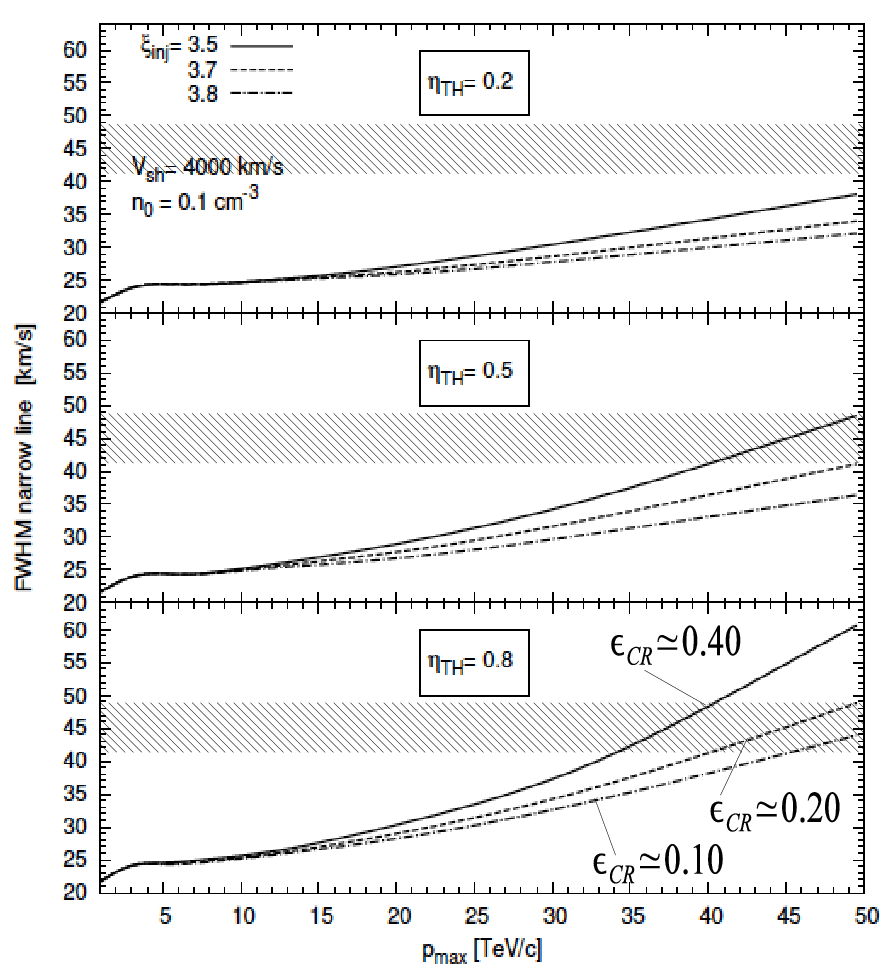}}
  \caption{FWHM of the narrow line as a function of the maximum momentum of accelerated protons. The three panels refer to $\eta_{\rm TH} =0.2, 0.5$ and 0.8. Each panel shows three lines calculated assuming different injection parameter, $\xi_{\rm inj} =3.5, 3.7$ and 3.8 which approximately correspond to $\epsilon_{CR} = 0.4, 0.2$ and 0.1. The shock speed is 4000 km s$^{-1}$, the total density 0.1 cm$^{-3}$ with 50\% of neutral fraction. The shadow band represent the FWHM measured from the {\it knot-g} in the Tycho's SNR.} 
 \label{fig:Tycho_narrow}
\end{center}
\end{figure}

2) The FWHM of the narrow component has been measured to be $44\pm4$ km s$^{-1}$ \cite{Sollerman03}. 
Such values implying a pre-shock temperature between 36,000 and 52,000 K. If this were the ISM equilibrium temperature there would be no atomic hydrogen, implying that the pre-shock hydrogen is heated by some form of shock precursor in a region that is  sufficiently thin so as to make collisional ionization equilibrium before the shock unfeasible. The CR precursor is the most plausible candidate to explain such a broadening of the narrow line width. To support this statement in Fig.~\ref{fig:Tycho_narrow} we compare the value measured from the {\it knot-g} in Tycho with the theoretical expectation calculated using the kinetic theory. 
The FWHM of the narrow Balmer line is plotted as a function of the maximum momentum $p_{\rm max}$ and for three different value od the turbulent heating $\eta_{\rm TH}=0.2, 0.5$ and 0.8 (from top to bottom panel). The three curves in each panel are obtained for $\xi_{\rm inj}=3.5,~3.7,~3.8$, which approximately correspond to $\epsilon_{\rm CR} = 0.4, 0.2$ and 0.1 for the chosed value of the shock velocity and total density, i.e. $V_{\rm sh}=4000\,\rm km\,s^{-1}$ and $n_{0}=0.1\,\rm cm^{-3}$.   The maximum momentum, $p_{\max}$ determines the spatial extent of the CR-induced precursor. Larger values of $p_{\rm max}$ imply that there is more time (space) for depositing heat in the upstream, and the width of the narrow Balmer line broadens correspondingly. The effect becomes more pronounced for larger values of the parameter $\eta_{\rm TH}$. 
The number of free parameters is too large and the FWHM of the narrow line alone cannot be used to constrain none of them. Nevertheless, the FWHM is an increasing function of all the three parameters, $p_{\max}$, $\eta_{\rm TH}$ and $\epsilon_{CR}$, hence if the shock speed, the upstream density and neutral fraction are known, we can put lower limit to these parameters. Following this procedure, in the case of Tycho we get: $p_{\max} \gtrsim 40$ TeV, $\eta_{\rm TH} \gtrsim 0.2$ and $\epsilon_{CR} \gtrsim 10\%$.
 
\section{Conclusions}
\label{sec:conclusion}
The quest for the origin of Galactic CRs remains open. The search for their origin is difficult and all new observational input need to be considered very carefully. Based mainly on the detection of gamma rays from several SNRs and on the morphology of the X-ray emission, which suggests strong magnetic field amplification, SNRs are considered, now more than ever before, as the most likely source of the bulk of CRs. Because the shocks produced by SN explosions often propagate in partially ionized gas, the emission lines of neutral hydrogen have recently been recognized as a possible diagnostic tool for CRs. There are three different signatures in the Balmer emission which could reveal the presence of CRs: 1) a shock that is accelerating particles is expected to be less effective in heating the background plasma, which is reflected in a Balmer-line emission with a smaller width than in the absence of CR acceleration; 2) the presence of a CR precursor upstream of the shock could heat the upstream plasma on a scale larger than the charge-exchange length but small than the ionization length, resulting in a larger width for the Balmer narrow line; 3) if the heating of the upstream plasma (especially electrons) is large enough, the collisional excitation between electrons and Hydrogen atoms can produce Balmer emission also from the region ahead of the shock.

Remarkably all these signatures has been observed in some SNRs. Here we summarized the results obtained by the kinetic theory of collisionless shocks in the presence of neutral hydrogen developed in \cite{BlasiMor12, MorBla12, MorBla13} comparing the theoretical predictions with data on the Balmer line emission from four SNRs: SN~1006, RCW 86, SNR 0509-67.5 and Tycho. While the NW region of SN~1006 is compatible with absence of CR acceleration, the other three remnants presents anomalies in the Balmer lies which can be explained assuming an acceleration efficiency around $\sim 10\%$ which is remarkably the amount of efficiency required for SNRs to be the main sources of Galactic CRs. Balmer emission could be used also to gather information on the CR-precursor length, on the magnetic damping in the precursor and  on the electron-to-proton temperature equilibration. 
Unfortunately the quality of data or the lack of important information (like the shock proper motion) prevent a better determination of the CR acceleration efficiency as well as other parameter involved in the theory. However the good news is that the amount and the quality of the data concerning the Balmer emission will increase soon in the next few years. For this reason the analysis of Balmer emission can be regarded as one of the most powerful technique to study the process of shock acceleration and, more generally, to study the physics of collisionless shock.

\section*{Acknowledgment}
I want to thank P. Balsi, E. Amato, R. Bandiera and D. Caprioli for our long-term collaboration on this topic and on cosmic ray physics in general. I am also deeply grateful to Nicoletta Mastroleo for her never-ending support and for sharing with me my doubts and uncertainties, solving many of them. This work has been partially founded through the NSF grant n.1306672, and it has been also  supported by a grant from the Simons Foundation and the hospitality of the Aspen Center for Physics during September 2013.
 



\nocite{*}
\bibliographystyle{elsarticle-num}
\bibliography{martin}



\end{document}